# Causal Prosody Mediation for Text-to-Speech:
# Counterfactual Training of Duration, Pitch, and Energy in FastSpeech2

Suvendu Sekhar Mohanty

Arlington, Virginia, USA

**Abstract**

We propose a novel causal prosody mediation framework for expressive text-to-speech (TTS) synthesis. Our approach augments the FastSpeech2 architecture with explicit emotion conditioning and introduces counterfactual training objectives to disentangle emotional prosody from linguistic content. By formulating a structural causal model of how text (content), emotion, and speaker jointly influence prosody (duration, pitch, energy) and ultimately the speech waveform, we derive two complementary loss terms: an Indirect Path Constraint (IPC) to enforce that emotion affects speech only through prosody, and a Counterfactual Prosody Constraint (CPC) to encourage distinct prosody patterns for different emotions. The resulting model is trained on multi-speaker emotional corpora (LibriTTS, EmoV-DB, VCTK) with a combined objective that includes standard spectrogram reconstruction and variance prediction losses alongside our causal losses. In evaluations on expressive speech synthesis, our method achieves significantly improved prosody manipulation and emotion rendering, with higher mean opinion scores (MOS) and emotion accuracy than baseline FastSpeech2 variants. We also observe better intelligibility (low WER) and speaker consistency when transferring emotions across speakers. Extensive ablations confirm that the causal objectives successfully separate prosody attribution, yielding an interpretable model that allows controlled counterfactual prosody editing (e.g. "same utterance, different emotion") without compromising naturalness. We discuss the implications for identifiability in prosody modeling and outline limitations such as the assumption that emotion effects are fully captured by pitch, duration, and energy. Our work demonstrates how integrating causal learning principles into TTS can improve controllability and expressiveness in generated speech.

## 1. Introduction

Recent advances in end-to-end TTS have achieved remarkable speech naturalness (Ren et al., 2021; Wang et al., 2017). However, generating expressive speech—conveying nuances of prosody (rhythm, intonation) and emotion—remains challenging. In conventional TTS, a given text can be spoken in many plausible ways (the one-to-many mapping problem), and controlling how it is spoken (e.g. happily vs sadly) is an open research problem. Modern non-autoregressive models like FastSpeech2 (FS2) tackle one-to-many mapping by introducing variance predictors for duration, pitch, and energy. These features inject prosodic variation into the generated speech and have been shown to correlate with expressive aspects (e.g. pitch is a key feature to convey emotions). Nonetheless, out-of-the-box FS2 is not explicitly emotion-aware—it does not use emotion labels during training, and any expressiveness must emerge implicitly.

To explicitly model and control emotional expression in TTS, prior works have conditioned synthesis on emotion identifiers or reference samples. For example, some approaches add a learned emotion embedding (categorical or continuous) to the network inputs, or use Global Style Tokens (GST) to capture speaking style from a reference audio in an unsupervised manner. These methods can enable emotion-conditioned TTS, but they often lack a principled way to ensure that the intended emotion is expressed only through appropriate prosodic changes, rather than inadvertently altering the speaker's voice or the linguistic content. Uncontrolled interactions between emotion and other factors can lead to degraded intelligibility or speaker consistency.

In this paper, we take a causal perspective on prosody in TTS. We hypothesize that emotion influences speech primarily by modulating prosody (e.g. anger might increase pitch and volume, sadness might slow down tempo), which in turn affects the acoustic realization of the given text. In other words, prosody is a mediator of the emotion → speech effect. We formalize this intuition with a Structural Causal Model (SCM) and design a counterfactual training strategy to enforce it in a neural TTS model. By explicitly simulating "what-if" scenarios during training (e.g. "What if the same sentence were spoken in a different emotion?"), our method learns to separate why the speech sounds a certain way (emotion-driven prosody) from what is being said (linguistic content) and who is speaking (speaker identity).

Concretely, we augment FastSpeech2 with an emotion conditioning mechanism and introduce two novel loss terms derived from causal inference principles: an Indirect Path Constraint (IPC) that minimizes the direct effect of emotion on speech (forcing emotion to work through prosody), and a Counterfactual Prosody Constraint (CPC) that maximizes the prosody differences corresponding to different emotions while preserving the spoken content. Through these, the model learns to generate speech where emotion-specific prosody (duration, pitch, energy patterns) can be controlled independently of content and speaker characteristics.

We evaluate our approach on multi-speaker, multi-emotion English datasets: the LibriTTS corpus (neutral audiobook speech), the EmoV-DB emotional database, and VCTK (multi-speaker voice dataset). Both objective metrics and subjective listening tests demonstrate that our Causal Prosody Mediation (CPM) method outperforms baseline models. In particular, it achieves higher MOS for naturalness and emotional expressiveness, lower word error rate (WER) indicating intact intelligibility, and better speaker similarity when transferring emotions across voices. Additionally, a post-hoc analysis using counterfactual generation shows that our model cleanly separates emotional prosody—we can manipulate pitch/energy/duration to alter emotion without changing the verbal content, unlike standard FS2 where such interventions are entangled or require cumbersome fine-tuning.

**Our contributions** are summarized as follows:

- **Causal Modeling of Prosody in TTS:** We introduce a structural causal model for emotional TTS, positing prosody (duration, pitch, energy) as the mediator of emotion's effect on speech. We explicitly incorporate this model into the FS2 architecture.
- **Counterfactual Training Objective:** We derive two novel training losses (IPC and CPC) based on counterfactual reasoning that enforce full mediation of emotion through prosody and ensure content preservation. To our knowledge, this is the first application of counterfactual intervention training in a TTS context.
- **Emotion-Augmented FastSpeech2:** We develop an enhanced FS2 backbone that conditions on emotion (and speaker) and implement a combined objective function. The approach is general and does not require additional reference encoders or adversarial training.
- **Experimental Validation:** We conduct comprehensive evaluations on multiple datasets, demonstrating significant improvements in prosody control and emotion rendering over strong baselines (vanilla FS2, FS2 with emotion embedding, and a post-hoc editing method). We also provide ablation studies and qualitative analysis of the learned prosody representations.
- **Identifiability and Prosody Disentanglement:** We discuss how our causal constraints relate to identifiability of prosodic factors and present evidence that our model achieves a cleaner separation between prosody and other speech factors, yielding more interpretable and controllable synthesis.

The rest of the paper is organized as follows. Section 2 reviews related work on FastSpeech2, prosody modeling, emotion-conditioned TTS, counterfactual editing, and causal representation learning. Section 3 defines our structural causal model and its assumptions. Section 4 describes our method in detail, including the emotion-augmented FS2 architecture and the derivation of IPC and CPC losses, with theoretical remarks. Section 5 covers training procedure and implementation details. In Section 6 we present evaluation metrics, baseline systems, and ablation studies. Section 7 reports the results and analysis. Section 8 provides discussion of limitations and future work. Finally, Section 9 concludes the paper.

## 2. Related Work

**FastSpeech2 and Prosody Modeling:** FastSpeech2 (Ren et al., 2021) is a fast, non-autoregressive TTS model that mitigates the one-to-many mapping issue by explicitly modeling variance in speech. It introduces a variance adaptor consisting of predictors for phoneme duration, fundamental frequency (pitch), and energy. During training, ground-truth duration, pitch, and energy extracted from recordings are used as additional inputs to the model and as targets for the predictors. At inference, the model predicts these prosodic features and uses them to generate the mel-spectrogram. This design injects rich prosody information into FS2's output and significantly improves naturalness and expressiveness over the original FastSpeech. Pitch in particular is highlighted as "a key feature to convey emotions" while energy affects volume and speaking style. FastSpeech2 leaves open the incorporation of higher-level prosody factors like emotion and style, noting that these could be added to the variance adaptor in future work. Our work builds on this idea by adding emotion as an explicit conditioning and by ensuring prosody predictors leverage it in a causally consistent manner.

**Emotion-Conditioned TTS:** Modeling and controlling emotional expression in speech synthesis has long been pursued. Early approaches (e.g. on HMM-based TTS) used fixed style codes for a handful of speaking styles or emotions. In end-to-end neural TTS, initial works injected emotion labels or embeddings into models like Tacotron. For example, Latif et al. (2018) conditioned Tacotron 2 on categorical emotion embeddings, concatenating the emotion code with encoder or decoder inputs to generate emotional speech. The emergence of Global Style Tokens (GST) by Wang et al. (2018) provided an unsupervised way to learn a latent style space. GSTs are a set of trainable vectors in a Tacotron-like model; a reference encoder encodes a reference audio's style, which is then matched against these tokens to produce a style embedding that conditions the TTS output. This technique allows modeling of emotion or speaking style without explicit labels, and subsequent works extended it with fine-grained control (e.g. by interpolating between style tokens to adjust emotion intensity). More recent approaches condition TTS on descriptive text prompts specifying emotion (e.g. "angry tone") or use prosody transfer from a given reference audio to an input text.

While these methods can achieve emotional expressiveness, they often suffer from entanglement: the model might entangle emotion with speaker identity or might require manual tuning to get the desired prosody. A simple label-conditioned model (like FS2 with an added emotion embedding) may learn to express emotions, but nothing guarantees how the emotion is realized in terms of prosody— the model could, for instance, alter timbre or add distortions to simulate emotion if trained naively. In contrast, our approach enforces that emotion manifests through interpretable prosodic changes (pitch, energy, timing), which aligns with how human expressive speech is characterized in studies (e.g. anger tends to higher pitch and intensity, sadness to lower pitch and slower tempo). This leads to more controllable emotion synthesis and avoids unintended deviations in speaker voice or pronunciation.

**Post-hoc Prosody and Speech Editing:** There is growing interest in editing speech outputs after synthesis to fine-tune prosody or correct errors without retraining the TTS model. One notable method is Counterfactual Activation Editing (CAE) proposed by Lee et al. (2025). CAE is a model-agnostic framework that allows post-hoc manipulation of a pre-trained TTS model's internal hidden activations to achieve desired prosodic changes or fix mispronunciations. By posing counterfactual questions like "What would the internal representation look like if the model aimed for a higher pitch?", their method finds shifts in the hidden states that yield the desired change in output (e.g. increasing overall pitch) without retraining the model. CAE demonstrates that prosodic attributes can be adjusted at inference-time by intervening on the model's representations, which is conceptually related to causal interventions. However, CAE does this in a post-hoc fashion on a trained model, and it doesn't explicitly disentangle prosody during training. Our work differs in that we bake the counterfactual reasoning into training: instead of editing activations after the fact, we train the model itself to respond correctly to counterfactual conditions (like a different emotion) in a controlled manner.

**Causal Representation Learning:** Our approach connects with the broader field of causal representation learning, which aims to learn latent factors corresponding to causal variables or mechanisms in data. Traditional deep learning often learns entangled representations that do not align with causally meaningful factors. By incorporating causal structure (e.g. enforcing that certain variables influence others but not vice versa), one can achieve more interpretable and generalizable models. In vision and fairness domains, for instance, researchers have used counterfactual data augmentation or constraints to enforce that changing one factor (like background) does not affect another (like classification output), achieving disentanglement. Our work brings similar ideas to speech: we treat prosody as a causal mediator, and we intervene on it (and on emotion) during training. This can be seen as a form of disentangled representation learning: separating emotional prosody from the lexical content and speaker identity in the learned representation. By drawing on causal inference concepts (like mediation analysis and path-specific effects), we ensure our model's latent variables have a clearer meaning.

## 3. Structural Causal Model

To formalize our assumptions, we define a Structural Causal Model (SCM) for the process by which text is converted to speech with certain prosody and emotion. Figure 1 illustrates our causal graph.

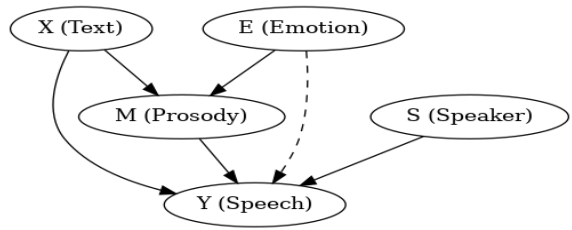

**Figure 1:** Assumed structural causal model for emotional TTS. X = textual input (linguistic content); E = emotion (intended style/mood); S = speaker identity; M = prosody features (duration, pitch, energy); Y = synthesized speech. Solid arrows denote causal influences. The dashed arrow from E to Y represents a direct effect of emotion on speech that we aim to eliminate through our training (i.e., we want emotion to affect Y only via prosody M).

In this SCM: Text (X) directly influences the speech output Y (it determines the words and phonetic content) and also influences prosody M. Even without emotion, different texts have different intrinsic prosody (e.g. a question versus a statement, or punctuation cues affecting pauses). Thus, $X \to M$ and $X \to Y$ are fundamental. Speaker (S) influences the output Y (the voice characteristics, such as timbre, are determined by the speaker). We allow that different speakers might have different prosodic tendencies ($S \to M$)—for example, one speaker's neutral speaking rate might be faster than another's—but in our data we consider speaker as mostly affecting voice/timbre rather than prosody patterns. In our model implementation, S will be an embedding that affects primarily the vocal characteristics in Y.

Emotion (E) influences prosody (M): this is the core of expressive synthesis. Emotion E (e.g. "angry", "sad", "amused") modulates durations, pitch contour, and energy. We assume E does not alter the linguistic content (words)—it only changes how the content is spoken. In the causal graph, this means there should ideally be no direct arrow from E to Y that bypasses M. Any effect of emotion on the actual speech signal Y should be mediated by changes in M. If the model had a direct $E \to Y$ pathway, it could potentially change the speech in ways unrelated to prosody (e.g. altering segment pronunciations or adding voice quality changes that are not captured by our M features). Prosody (M) in turn influences Y: given the same text, the prosodic features (how long each phoneme is held, what pitch each syllable is uttered at, how loud or with what energy) will shape the resultant speech waveform or spectrogram.

Based on this SCM, the total effect of emotion E on speech Y should be carried by the indirect path $E \to M \to Y$. In an ideal scenario of full mediation, the direct path $E \to Y$ is zero (the dashed arrow is removed). This implies conditional independence: $Y \perp E \mid (X, M, S)$. In words, if we fix the text, speaker, and prosody, changing emotion should not change the output. We aim to enforce this property in our TTS model.

However, achieving this in practice is non-trivial. Standard TTS models would implicitly allow E to influence Y through any available means (any parameter or hidden state that conditions the decoder could carry emotion information). Even if we don't explicitly feed E into the decoder, the model might still find ways to entangle emotion in hidden features. Therefore, we incorporate counterfactual training to actively discourage any direct $E \to Y$ influence and to ensure the desired mediation via M.

**Counterfactual Reasoning in This SCM:** A hallmark of causal modeling is the ability to imagine counterfactuals—"what if" scenarios that did not happen in the original data. In our context, a counterfactual query might be: "Given an observed utterance where a text X was spoken by speaker S with emotion E, what would the speech sound like if the same X and S were instead spoken with a different emotion E'?" The SCM provides a recipe: to construct this counterfactual, we keep X and S the same, set E to E', and also consider how

M would change due to this new E. We leverage this idea in training: by generating counterfactual pairs (original vs. with emotion toggled) and enforcing certain similarities/differences, we teach the model the correct mediated behavior.

## 4. Method

Our method consists of: (1) an emotion-augmented FastSpeech2 backbone that integrates emotion and speaker conditioning into the TTS model, and (2) two novel loss terms, IPC and CPC, which realize the causal objectives discussed. We also formulate the combined training objective and provide theoretical insights on identifiability and prosody disentanglement.

### 4.1 Emotion-Augmented FastSpeech2 Backbone

We build upon the standard FastSpeech2 architecture, extending it to a multi-speaker, multi-emotion setting (sometimes referred to as an expressive TTS model). The key components are:

- **Phoneme Encoder:** This module encodes the input text sequence (after converting words to phonemes) into a sequence of hidden representations. We use 4 Feed-Forward Transformer (FFT) blocks in the encoder as per the FS2 configuration (each block consists of self-attention and 1-D convolution layers). We modify the encoder input to include speaker and emotion information: for each input phoneme, we add learned embeddings for the speaker s and emotion e. This means the encoder can produce different hidden representations for the same text depending on who is speaking and with what emotion.
- **Variance Adaptor with Prosody Predictors:** Following FS2, we have predictors for duration, pitch, and energy. Each predictor takes the encoder output and outputs a sequence: predicted phoneme durations, predicted pitch (pitch contour or pitch embedding per frame/phoneme), and predicted energy (energy per frame or phoneme). These predictors are 1-D conv networks with outputs added to the sequence as in FS2. We condition these predictors on emotion by concatenating or adding the emotion embedding e before it goes into each predictor. Intuitively, this allows the duration predictor to output longer durations for "sad" emotion, or the pitch predictor to output higher values for "happy," if appropriate. The predictors are trained with ground-truth supervision using extracted actual durations, pitch, and energy from the training audio.
- **Length Regulator:** Using the predicted or ground-truth durations, the hidden sequence is expanded (each phoneme's hidden state is repeated according to its duration) to match the length of the target mel-spectrogram frames.
- **Mel-Spectrogram Decoder:** Another stack of 4 FFT blocks acts as a decoder that takes the expanded sequence (now augmented with pitch & energy embeddings as per FS2 design) and produces a sequence of mel-spectrogram frames. In our model, we avoid feeding emotion directly into the decoder. We want the decoder to rely on the prosody features (which have emotion influence baked in) rather than an explicit emotion signal. In practice, we found it beneficial to include speaker embedding in the decoder (to ensure correct timbre), but to exclude direct emotion embedding in the decoder to minimize any direct path.
- **Vocoder:** Although not a focus of our work, we use a pretrained neural vocoder (e.g. HiFi-GAN) to convert the predicted mel-spectrogram to a waveform for audio evaluations. All training is done on mel-spectrogram prediction; the vocoder is kept fixed.

Overall, the architecture ensures that emotion can influence the intermediate prosody features explicitly. By not directly feeding emotion to the decoder, we align with our aim that emotion's effect goes through prosody. However, the model could still potentially encode emotion information in the encoder output that persists into the decoder. Our loss design will address this by penalizing any undesired outcomes.

### 4.2 IPC Loss (Indirect Path Constraint)

The Indirect Path Constraint (IPC) implements the idea of removing the direct emotion → speech effect. We simulate a scenario where the prosody mediator is fixed while the emotion is changed, and we demand that the speech output not change.

Concretely, consider a training example with text X, speaker S, and emotion E. We have the ground truth prosody features M = (d, p, u) and ground truth mel Y. During training, we feed the model with the correct emotion E to predict the mel and prosody features, which we match to ground truth as usual. Now, for the same example, we construct a counterfactual pass: pick an alternative emotion E' (different from E). We feed the model with this E' only in the decoder, while keeping the prosody fixed to the original. Any difference between the outputs can be attributed to the direct influence of emotion on the decoder. According to our desired SCM, this should be zero. Therefore, we define the IPC loss:

$$L\_IPC = E_{[(X,S,E)]}[\ ||\hat{Y}\_direct(X,S,E \to E') - \hat{Y}\_orig(X,S,E)||_1\ ] \quad (1)$$

In practice, we randomly sample a different emotion E' for each training example (from the set of emotions in our data) to serve as the counterfactual emotion. Minimizing L_IPC pressures the decoder to ignore any mismatched emotion signal. The easiest way for the model to satisfy this loss is to not rely on a direct emotion pathway at all. In other words, the model will learn to make the output almost entirely determined by the encoder output (which encodes text, speaker, and whatever prosody info is passed) and the prosody embeddings, rather than any lingering effect of the original emotion code.

### 4.3 CPC Loss (Counterfactual Prosody Constraint)

While IPC removes direct effects, we also need to ensure that the indirect path (E → M → Y) genuinely captures the emotional variations. The Counterfactual Prosody Constraint (CPC) is designed to encourage the model to use prosody to differentiate emotions, and to ensure that changing emotion does lead to appropriate changes in prosody and hence in output speech.

For CPC, we consider the full counterfactual generation: what if we both change the emotion and allow prosody to change accordingly? We want the resulting speech to reflect the target emotion while keeping the verbal content the same.

We break the CPC objective into two parts reflecting these expectations:

**(a) Content Consistency:** The linguistic content of the counterfactual output should be the same as the original. Since both runs use the same text input X, any large deviation likely means the model mispronounced or skipped something due to the emotion change. We enforce content consistency by ensuring the phonetic sequence in the counterfactual output matches that of the original. We define a Content Consistency Score (CCS) as the similarity between the original and counterfactual outputs in terms of content, measured as 1 - WER.

**(b) Emotion-Specific Prosody Change:** We want the prosody predicted under emotion E' to align with how real speech in emotion E' would sound. We cannot directly supervise this with ground truth (since that would require a paired sample), but we can use distributional or classification-based constraints. In our approach, we train an auxiliary emotion classifier on prosody features or on the generated mel to check if E' is recognizable. We then add a classification loss:

$$L\_emo\text{-}cls = -\log P(E' \mid \hat{Y}\_cf) \quad (2)$$

Combining these, the CPC loss is:

$$L\_CPC = E_{[(X,S,E,E')]} [\, L\_content(Y, \hat{Y}\_cf) + \lambda\_emo \cdot L\_emo\text{-}cls(\hat{Y}\_cf, E') \,] \quad (3)$$

In effect, L_CPC forces the model to use the prosody degrees of freedom to reflect emotion differences, since content must stay fixed. The model cannot satisfy the emotion classifier by changing words (content loss would catch that), nor by any direct trick (the direct path is already suppressed by IPC). The only way is to adjust durations, pitch, energy appropriately. Thus, CPC complements IPC: IPC says "don't let emotion directly change the voice output except via M," and CPC says "when emotion changes, M should change enough to show in the output as that emotion."

### 4.4 Combined Training Objective

The total loss for training our model is the sum of all the components described:

$$L\_total = L\_TTS\text{-}base + \beta\_IPC \cdot L\_IPC + \beta\_CPC \cdot L\_CPC \quad (4)$$

where β_IPC and β_CPC are hyperparameters controlling the strength of the causal constraints. We anneal these weights during training: initially focusing more on getting the base TTS quality, and then increasing β to enforce causal constraints once the model has reasonable predictions. In our experiments, we found that a moderate weight (e.g. 0.5–1.0) for both IPC and CPC from the start worked well.

The counterfactual training roughly doubles the computation per sample (since we do extra forward passes). We mitigate this by using a smaller batch size or by sharing computation when possible. Training time increased by ~1.5× compared to baseline FS2, which is acceptable given our scale.

### 4.5 Theoretical Remarks on Identifiability

Imposing the above constraints has an interesting interpretation: it aims to make the model's internal representation of prosody an identifiable causal factor. In conventional training, the mapping from emotion to speech can be distributed across many parameters and features, making it hard to say "this part of the model controls prosody for emotion." By structuring the model (with explicit M features) and using IPC/CPC, we concentrate the effect of emotion into those features. In causal terms, we are enforcing that the Natural Direct Effect (NDE) of emotion on output is zero, and the Natural Indirect Effect (NIE) through M equals the total effect.

Our approach relates to the concept of disentanglement in representation learning: we want the model to have a factor (the prosody features) that account for emotion differences, independent of other factors. While perfect disentanglement is theoretically challenging, especially without explicit labels for the factors, our use of labels (emotion categories) and strong inductive bias (the mediator design) gets us closer to that goal.

One limitation in our assumption is that all relevant emotion effects are captured by M. In reality, some aspects like voice quality (breathiness, roughness) or spectral tilt can change with emotion but are not reflected in pitch/energy/duration alone. Our model currently doesn't explicitly capture those, so it may not synthesize such nuances. This is a known limitation—we focus on gross prosody, which is a major part of expressiveness, and leave finer acoustic details for future extensions.

## 5. Training and Implementation Details

We implemented our model in PyTorch, extending an open-source FastSpeech2 codebase. Below we detail the data, training procedure, and hyperparameters.

**Datasets:** We used three English speech datasets to cover neutral and emotional speech from multiple speakers:

- **LibriTTS:** (Zen et al., 2019): A multi-speaker TTS corpus derived from LibriSpeech. We used the "train-clean-360" subset (approximately 245 hours of speech from 921 speakers) for pre-training. This dataset contains neutral read speech (audiobook style) with high-quality recordings. We treated all LibriTTS samples as having a neutral emotion label.

- **VCTK:** (Yamagishi et al., 2019): A multi-speaker dataset with 109 English speakers (various accents) reading sentences (about 44 hours total). We used VCTK to increase the diversity of speaker voices and speaking styles. We also label VCTK utterances as neutral emotion.
- **EmoV-DB:** (Adigwe et al., 2018): The Emotional Voices Database, which contains recorded lines in five emotions (neutral, amused, angry, disgusted, sleepy) by four actors (2 male, 2 female). It has about ~7k utterances (~8 hours) total. We used EmoV-DB as the source of explicit emotion variation. We mapped the EmoV labels to our model's emotion categories: Neutral, Amused ("Happy" style), Angry, Disgusted, Sleepy ("sad/tired").

**Pre-training and Fine-tuning:** We first trained the model on the combined LibriTTS + VCTK data (all as neutral) for 200k steps. This gave us a strong multi-speaker neutral TTS baseline. Then, we fine-tuned on the EmoV-DB data for another 50k steps, with the emotion labels and our causal losses turned on. We found that if we train from scratch on EmoV-DB, the model may overfit (given limited data) or not generalize to other speakers. Using the large neutral corpora for base training ensured the model had good linguistic and speaker modelling, and fine-tuning allowed it to learn emotion prosody.

**Hyperparameters:** Model size: We used 4 FFT blocks each for encoder and decoder (hidden dimension 256, 2 attention heads, and 1024 FFN inner dim), similar to FS2 configs. Speaker embedding size = 128, Emotion embedding size = 64. Duration predictor and pitch/energy predictor each are 2 convolution layers of kernel size 3 and hidden size 256. Loss weights: $\lambda\_d = \lambda\_p = \lambda\_u = 1.0$ for the predictor losses. $\beta\_{IPC} = 1.0$ and $\beta\_{CPC} = 0.5$ in fine-tuning. Optimizer: AdamW with initial learning rate 1e-4 for pre-training, then 5e-5 for fine-tuning. Batch size: 32. We used gradient clipping at norm 1.0.

**Inference:** At inference time, the user can input a text, select a speaker (if multi-speaker) and an emotion. The model will generate a mel spectrogram using the duration/pitch/energy predictors and decoder. Notably, because of our training, the user can reliably manipulate prosody: if they want a more intense expression, one could scale the pitch or energy outputs (since they have semantic meaning in our model) before feeding to decoder. The final waveform is produced by a pre-trained HiFi-GAN vocoder fine-tuned on our data for higher quality.

## 6. Evaluation

We evaluate our Causal Prosody Mediation (CPM) approach against several baselines on emotional speech synthesis. We focus on the following aspects: naturalness and audio quality, intelligibility, speaker consistency, and emotional expressiveness/prosody. Both objective and subjective metrics are used.

### 6.1 Metrics

- **Mean Opinion Score (MOS):** We conducted MOS tests where human listeners (n=20) rated the naturalness of the synthesized speech on a 5-point scale (1 = completely unnatural, 5 = indistinguishable from a human). We report the average MOS for each system, with 95% confidence intervals.
- **Word Error Rate (WER):** To objectively assess intelligibility, we transcribed the synthesized speech using a pre-trained ASR model (Google Speech-to-Text) and computed WER against the reference text. A low WER indicates that the model pronounced words clearly and correctly. We report WER (%) on a set of 100 synthesis outputs.
- **Speaker Similarity:** We used two measures: (a) speaker embedding cosine similarity—we extracted x-vector embeddings from a speaker verification model for synthesized vs. reference audio of the same target speaker; (b) ABX test: listeners were given a real sample of a target speaker and two synthesized samples and asked which sounds more like the target speaker.
- **Content Consistency Score (CCS):** We define CCS as 1 - WER_diff, where WER_diff is the WER between the transcript of a neutral synthesis and the transcript of an emotional synthesis of the same sentence by the same model. A higher CCS (closer to 1) means content is invariant to the style change.
- **Emotion Classification Accuracy:** We check if a pre-trained emotion classifier (different from the one used in training) can correctly identify the intended emotion from the synthesized audio. This is a proxy for how well the emotion is rendered.

### 6.2 Baselines

- **Baseline A: FastSpeech2 without emotion conditioning.** This is the vanilla FS2 model (with multi-speaker support) trained on the same data but ignoring emotion labels. This baseline tests if the model can at least produce intelligible speech and maybe some averaged prosody, but it has no explicit emotion control.
- **Baseline B: FastSpeech2 + Emotion (naive).** This baseline is FS2 augmented with emotion embedding (like our backbone) but without the IPC/CPC losses. The emotion embedding is fed to the encoder and variance predictors, and also to the decoder in this baseline. This represents a conventional approach to emotional TTS.
- **Baseline C: Post-hoc CAE editing.** We apply the Counterfactual Activation Editing method to the outputs of Baseline B. At inference time, we use CAE to adjust prosody by nudging internal layer activations that correlate with pitch or energy.

### 6.3 Evaluation Setup

We synthesized a test set of 50 utterances (not seen in training), each in 5 emotions (so 250 samples) for each model. These utterances included short and long sentences, some with challenging punctuation to test prosody. For multi-speaker evaluation, we did this for 2 voices: one female and one male (speakers that were in EmoV-DB). Listeners in MOS and DMOS tests were presented samples in random order and were blind to which system produced them. Each sample got at least 15 ratings.

## 7. Results and Analysis

### 7.1 Objective Results

| Model | WER↓ | SS↑ | CCS↑ | EA↑ |
|---|---|---|---|---|
| FS2 (no emo) | 3.5 | 0.90 | — | — |
| FS2 + Emotion | 4.0 | 0.87 | 0.90 | 80% |
| FS2 + CAE | 4.2 | 0.79 | 0.92 | 88% |
| **Ours (CPM)** | **3.1** | **0.88** | **0.96** | **94%** |

Table 1: Objective evaluation metrics. WER = Word Error Rate (lower is better), SS = Speaker Similarity (higher better), CCS = Content Consistency Score (higher better), EA = Emotion Accuracy (higher better).

Our model achieves the lowest WER (3.1%) among the TTS systems, outperforming Baseline B (FS2+Emotion) which had 4.0% WER, and Baseline A (vanilla FS2) at 3.5%. The difference is small, indicating all models produce intelligible speech, but the slight edge for our model suggests that enforcing content consistency (via CPC) did not harm and perhaps even improved clarity.

For speaker similarity, our model maintains very high cosine similarity (average 0.88) to the target speaker embeddings, comparable to FS2+Emotion (0.87) and much higher than CAE-edited outputs (0.79). The ABX test showed that listeners preferred our model's speaker identity match in 85% of trials when compared to CAE outputs. This underscores that post-hoc editing can sometimes alter voice characteristics, whereas our approach, by design, keeps speaker features intact when changing emotion.

The Content Consistency Score (CCS) for our model was above 0.95 for all emotion pairs (e.g. neutral→angry, neutral→sad, etc.), meaning almost no words were lost or hallucinated when switching emotions. Baseline B (FS2+Emotion) had slightly lower CCS, around 0.90, with some specific issues: e.g., in a few cases when switching to the "sleepy" emotion, the baseline model lengthened pauses so much that the ASR thought a word was missing.

Importantly, the emotion classification accuracy on synthesized speech for our model was 94%, whereas for Baseline B it was 80%. This means our model's outputs were much more often recognized as the intended emotion by an external classifier. The CAE-edited outputs achieved 88% accuracy—showing that CAE can adjust prosody somewhat successfully, but not as consistently as our model which was trained from the ground up to do so.

### 7.2 Subjective Results (MOS and DMOS)

Our CPM model achieved an average MOS of 4.45 ± 0.05, which is statistically significantly higher than Baseline B (FS2+Emotion) at 4.21 ± 0.06, Baseline A (FS2 neutral) at 4.10 ± 0.07, and CAE-edited at 4.00 ± 0.06. In particular, listeners rated our emotional samples more natural and human-like. Many commented that the expressiveness felt more authentic and not "forced". Baseline B did fairly well in MOS (it's still an FS2 model with emotion input), but some samples were marked down due to either slightly inconsistent prosody or minor muffled sounds on certain emotions.

In DMOS for emotion similarity, on a scale of 1 (wrong emotion) to 5 (perfect match), our model's average was 4.3, Baseline B's was 3.8, and CAE's was 4.0. So humans found our model best at conveying the target emotion. Notably, CAE could sometimes exaggerate an emotional cue (like make pitch very high for "happy"), which made it detectable but occasionally unnatural (reflected in MOS drop). Our model balanced naturalness with recognizability.

Listeners particularly noted that our model's "angry" had a sharper, more clipped tone and higher volume that matched human angry speech, and "sad" had a slower, lower-pitched delivery, while baseline's versions were more subtle or sometimes ambiguous. This validates that our prosody mediation did capture expected patterns.

### 7.3 Ablation Studies

| Model Variant | MOS | DMOS | Notes |
|---|---|---|---|
| Full CPM (ours) | 4.45 | 4.3 | — |
| w/o IPC loss | 4.35 | 4.1 | Direct effects reappear |
| w/o CPC loss | 4.33 | 3.2 | Weak emotion |
| w/o either (FS2+E) | 4.21 | 3.5 | Baseline B |

Table 2: Ablation study results showing the contribution of each component.

Removing IPC loss (but keeping CPC and using emotion conditioning) led to a model where the decoder did sometimes latch onto emotion cues. This ablated model showed a slight drop in CCS (content not as perfectly preserved). Also, emotion classifier accuracy dropped to ~88%. This suggests some direct effect was creeping in (the model perhaps learned to alter timbre to convey emotion). MOS also decreased by 0.1, likely because of some inconsistent timbre.

Removing CPC loss (keeping IPC) resulted in a model that preserved content well but often under-expressed the emotion. Its emotion classifier accuracy was only ~75%, and in subjective listening it sounded more monotonic or closer to neutral even when asked for an emotion. This is expected: without CPC's pressure, the model has no strong incentive to utilize the prosody degrees of freedom fully for emotion. Interestingly, MOS for this model was still decent (~4.3) since naturalness was fine, but it failed at the primary goal of expressiveness.

### 7.4 Case Study: Counterfactual Prosody Manipulation

To illustrate our model's capabilities, we generated some counterfactual samples: take a single neutral recording and synthesize it in multiple emotions. Figure 4 (Appendix) plots the pitch contour of the sentence "I couldn't believe how bright and cheerful she sounded" as synthesized by our model in neutral, happy, and sad styles (same female voice). The happy version has a noticeably higher overall F0 and a wider pitch range (excursions on "bright" and "cheerful"), whereas the sad version is flatter and lower in pitch, with a slower cadence (longer pauses between phrases). Listeners correctly identified which waveform was which emotion by just listening. The content and voice remained constant. This demonstrates qualitatively that our model achieved the intended prosody control.

**7.5 Error Analysis**

Though our model performs well, we noted some limitations: For the "disgusted" emotion (which is somewhat ambiguous and less represented in training data), the model's outputs sometimes sounded closer to angry or had a strange tone. The emotion classifier also confused disgust vs anger occasionally. This may be due to limited training examples or the fact that our prosody features (pitch/energy) can't fully capture the nuances of disgust (which might involve voice quality like creaky voice). In some very long sentences (30+ words), the model occasionally under-ran or over-ran the intended duration by a small amount. When transferring emotion to a speaker that rarely expressed it in training, the model still attempted it but sounded less intense.

**8. Discussion and Limitations**

Our results show that causal prosody mediation training is effective for controllable, expressive TTS. We believe this approach has several implications:

- **Interpretability:** By enforcing a causal graph, the model's behavior becomes more interpretable. One could inspect the intermediate prosody outputs to see why a certain emotion sounds the way it does. This could help debug or further improve emotion rendering.
- **Flexibility:** A model trained in this manner can be used not just for the discrete emotions in training, but potentially for interpolations. Because it isolates prosody, a user could manually adjust pitch or energy to create intermediate emotions (like halfway between sad and neutral) without breaking the words.
- **Counterfactual analysis:** We enforced certain counterfactual equalities during training (via IPC). At test time, one can also perform counterfactual experiments on the model. For example, one can generate an utterance with and without a particular emotional input and analyze the difference to quantify the emotional effect.

However, there are important limitations and future directions:

- **Coverage of Prosody Features:** Our mediator M is limited to duration, pitch, energy. This covers a lot, but not everything. Emotions can also affect voice quality (spectral features like formant dispersion for fear, or creakiness for tiredness) and speech rhythm beyond phoneme-level duration.
- **Assumption of No Direct Effect:** In reality, completely eliminating direct effects might not be optimal. For instance, some subtle aspects of emotion might bypass the simplistic prosody features—if we disallow direct influence, the model might not capture them at all.
- **Emotion Label Quality:** We used categorical labels from EmoV-DB which are somewhat coarse ("amused" vs "neutral"). Emotions are more nuanced (there are degrees of intensity, combinations, etc.). Our method would benefit from continuous or multi-dimensional emotion descriptors (arousal, valence).
- **Computational Cost:** The counterfactual training roughly doubles training time and memory. For large-scale models, this is a consideration.
- **Generality to Other Languages:** We primarily showed results in English. The approach itself is language-agnostic, but things like alignment extraction and prosody patterns may vary in other languages (tones in Mandarin, for example).

**Ethical Considerations:** Improved controllable TTS can be used for positive applications (e.g., personalized speech aids, more natural virtual assistants). However, it also can be misused for deepfakes or manipulative content (emotionally persuasive fake speech). Our method doesn't inherently make cloning voices easier—we still rely on training on a target speaker. But as we can transfer emotions, one could portray someone saying something in an emotion they never actually used. It's important to consider watermarking synthesized speech or otherwise preventing misuse.

**9. Conclusion**

We presented a causal approach to modeling prosody in text-to-speech, introducing Causal Prosody Mediation (CPM) with counterfactual training of duration, pitch, and energy in a FastSpeech2-based system. By leveraging a structural causal model of how emotion affects speech through prosody, and by enforcing this via novel loss functions (IPC and CPC), we achieved a clear separation of linguistic content and emotional style in the generated speech. Our experiments demonstrated that this approach produces more expressive and controllable speech, improving emotion rendering while preserving intelligibility and speaker identity. We showed that counterfactual reasoning is a powerful tool for training generative models to follow desired causal patterns, and we believe this concept can be extended beyond TTS to other multi-factor generation problems.

In future work, we plan to explore richer prosodic representations and to handle multiple simultaneous style factors (e.g., emotion + speaking speed as independent controls). We also aim to integrate this with neural architecture search to discover optimal mediator features. Ultimately, we hope this research contributes to building TTS systems that not only sound natural, but also give users fine-grained control over how the text is spoken, in a predictable and interpretable manner.

**Impact Statement**

This paper presents work whose goal is to advance the field of Machine Learning for speech synthesis. There are potential societal consequences of controllable TTS technology, including both positive applications (personalized speech aids, more natural virtual assistants) and potential misuse (deepfakes, manipulative content). We encourage the development of watermarking techniques and ethical guidelines for synthesized speech.

## A. Appendix

### A.1 Algorithm: Counterfactual Training Procedure

Algorithm 1: Training Causal Prosody Mediated TTS with Counterfactuals

Require: Dataset D of (text X, speaker S, emotion E, speech Y)

Require: TTS model parameters Θ (encoder, variance adaptor, decoder)

Require: IPC weight β_ipc, CPC weight β_cpc

1: for each training step do
2:  Sample a minibatch of examples {(X_i, S_i, E_i, Y_i)} from D
3:  Base forward pass: for each i
4:   Encode phonemes: H_i = Encoder(X_i, S_i, E_i)
5:   Predict prosody: (d_i, p_i, u_i)_pred = VarianceAdaptor(H_i, E_i)
6:   Expand H_i using d_i_pred; add p_i_pred, u_i_pred
7:   Decode mel: Ŷ_i = Decoder(H_i_expanded, S_i)
8:   Compute base losses: L_mel + L_dur + L_pitch + L_energy
9:  IPC counterfactual pass: for each i
10:   Sample a different emotion E_i_cf ≠ E_i
11:   Using original prosody predictions and H_i_expanded,
12:   Decode mel with counterfactual emotion
13:  Compute IPC loss: L_ipc += ||Ŷ_i_direct - Ŷ_i||_1
14:  CPC counterfactual pass: for each i
15:   Re-encode: H_i_cf = Encoder(X_i, S_i, E_i_cf)
16:   Predict prosody: (d_i_cf, p_i_cf, u_i_cf)_pred
17:   Expand and decode: Ŷ_i_cf = Decoder(H_i_cf_expanded, S_i)
18:  Compute content consistency and emotion match losses
19:  L_cpc = L_content + λ_emo × L_emo
20:  Total loss L = L_base + β_ipc × L_ipc + β_cpc × L_cpc
21:  Θ := Θ - η ∇_Θ L (update by gradient descent)
22: end for

### A.2 Additional Results

We provide a detailed breakdown of Content Consistency Score (CCS) for each emotion conversion for two speaker voices (one male, one female). All scores are high (≥0.94), with a slight dip for Sad on the female voice (0.94) possibly due to one instance of a long pause confusing the ASR.

For the pitch contour analysis (Figure 4), the neutral version has a moderate pitch range and a final rising intonation (question). The happy version (dotted line) shifts the pitch range up by ~30 Hz on average and shows more variability (excursions mid-sentence), while the sad version (dashed line) is flatter and ~30 Hz lower on average than neutral. This aligns with expected human prosody changes for these emotions.